\documentclass[10pt,twoside]{article}
\usepackage[applemac]{inputenc}
\usepackage{amssymb,amsmath}
\def\volnum{Manuscrit}
\usepackage{aflb}
\pdebut{1}
\def\pacs#1{\LP P.A.C.S.: #1}
\title{Electroweak gauge fields, particles, 
and antiparticles arise from probability}
\titleshort{Electroweak gauge fields \dots}
\author{Gunn Quznetsov}
\authorshort{G. Quznetsov}
\address{454016, Chelyabinsk-16, yD.BET. $\Phi $ N 949892, Russia\\
lak@cgu.chel.su, gunn@mail.ru, gunn@chelcom.ru}

\begin{document}
\maketitle

\vskip 1cm
\begin{abstract}

R\'ESUM\'E. Particules, antiparticules et les champs de m\'emeque les 
\'electromagn\'etiques calibriques champs sont d\'eduit des probabilit\'es des 
\'el\'enement phisiques \`a condition que ses \'el\'enements sont pr\'eset\'es 
par les spinors.

{\it 
ABSTRACT. Probabilities of events are expressed by the spinor functions 
and by operators of a probability creation and by operators of a probability 
annihilation. The motion equations in form of the Dirac equations with the 
additional fields are obtained for these spinor functions. Some of these 
additional fields behave as the gauge fields and others behave as the mass 
members. Such motion equations, contained all five elements of Clifford's 
pentad, is invariant for the electroweak gauge transformations. The creation 
operators and the annihilation operators of particles and antiparticles 
obtained from these probabilistic functions. The motion equation of the U(2) 
Yang-Mills field components is formulated similarly to the Klein-Gordon 
equation with the nonzero mass.}
\end{abstract}
\pacs{02.50.Cw; 11.30; 11.40; 11.80.F; 11.10; 03.70}

\section{Introduction}

In this article I do not construct a model for the particles physics but there 
I try to find the Quantum Theory principles which can be logically obtained from 
probabilities of physical events under the expression of these probabilities by 
spinors. Almost all basic principles prove to be of such kind. 

I call an event, occurred at single point of the space-time, as {\it a point
event}.

In Part 2 of this article probabilities of point events are expressed by the 
spinor functions and by {\it operators of a probability creation} and {\it a 
probability} {\it annihilation}. These operators are similar to the field 
operators of Quantum Fields Theory. The motion equations in form of the Dirac 
equations with the additional fields are obtained for the spinor functions. 
Some of these additional fields behave as the mass members and others behave as the
gauge fields.

Further I use the following denotation:

\[
1_2\stackrel{def}{=}\left[ 
\begin{array}{cc}
1 & 0 \\ 
0 & 1
\end{array}
\right] \mbox{, }0_2\stackrel{def}{=}\left[ 
\begin{array}{cc}
0 & 0 \\ 
0 & 0
\end{array}
\right] \mbox{, } 
\]

the Pauli matrices:

\[
\sigma _1=\left( 
\begin{array}{cc}
0 & 1 \\ 
1 & 0
\end{array}
\right) \mbox{, }\sigma _2=\left( 
\begin{array}{cc}
0 & -\mathrm{i} \\ 
\mathrm{i} & 0
\end{array}
\right) \mbox{, }\sigma _3=\left( 
\begin{array}{cc}
1 & 0 \\ 
0 & -1
\end{array}
\right) \mbox{.} 
\]

A set $\widetilde{C}$ of complex $n\times n$ matrices is called as {\it %
Clifford's set} \cite{Md} if the following conditions are fulfilled:

if $\alpha _k\in \widetilde{C}$ and $\alpha _r\in \widetilde{C}$ then $%
\alpha _k\alpha _r+\alpha _r\alpha _k=2\delta _{k,r}$;

if $\alpha _k\alpha _r+\alpha _r\alpha _k=2\delta _{k,r}$ for all elements $%
\alpha _r$ of set $\widetilde{C}$ then $\alpha _k\in \widetilde{C}$.

If $n=4$ then Clifford's set either contains $3$ matrices ({\it %
Clifford's triplet}) or contains $5$ matrices ({\it Clifford's pentad}).

There exist only six Clifford's pentads\cite{Md}: one {\it light pentad} $%
\beta $:

\begin{equation}
\beta ^{\left[ 1\right] }\stackrel{def}{=}\left[ 
\begin{array}{cc}
\sigma _1 & 0_2 \\ 
0_2 & -\sigma _1
\end{array}
\right] \mbox{, }\beta ^{\left[ 2\right] }\stackrel{def}{=}\left[ 
\begin{array}{cc}
\sigma _2 & 0_2 \\ 
0_2 & -\sigma _2
\end{array}
\right] \mbox{, }\beta ^{\left[ 3\right] }\stackrel{def}{=}\left[ 
\begin{array}{cc}
\sigma _3 & 0_2 \\ 
0_2 & -\sigma _3
\end{array}
\right] \mbox{,}  \label{lghr}
\end{equation}

\begin{equation}
\gamma ^{\left[ 0\right] }\stackrel{def}{=}\left[ 
\begin{array}{cc}
0_2 & 1_2 \\ 
1_2 & 0_2
\end{array}
\right] \mbox{, }  \label{lghr1}
\end{equation}

\begin{equation}
\beta ^{\left[ 4\right] }\stackrel{def}{=}\mathrm{i}\cdot \left[ 
\begin{array}{cc}
0_2 & 1_2 \\ 
-1_2 & 0_2
\end{array}
\right] \mbox{;}  \label{lghr2}
\end{equation}

three {\it chromatic} pentads and two {\it taste} pentads \cite{Q42}.

In this paper I consider the motion equations such that these equations
contain only the light pentad elements. Such equations are called as the{\it \ }%
equations for {\it a leptonn }motion \footnote{%
I use the terms "leptonn", "bosonn" etc with double "nn" for the
distinguishing of the physical notions and the logically receiving notions.}.

The Dirac equation contains only four elements of Clifford's pentads. Three
of these elements (\ref{lghr}) conform to three space coordinates and the
fourth element (\ref{lghr1}) either constitutes the mass member or conforms
to the time coordinate. But Clifford's pentad contains five elements.
Certainly, the fifth element (\ref{lghr2}) of the pentad should be added to
the motion equation. Hence the Dirac equation mass part will hold two
members. Moreover, if two additional quasi-space coordinates \cite{D5} will
be put in accordance to these two Clifford's pentads mass elements ((\ref
{lghr1}) and (\ref{lghr2})) then the homogeheous Dirac equation will be
obtained. All the five elements of Clifford's pentads and all the five space
coordinates are equal in quality in this equation. The magnitudes of all the
local velocities are equal to unit at such five- dimensional space.

The redefined in this way motion equation is invariant for rotations at the
2-space of the fourth and the fifth coordinates. In Part 3 this transformation 
defines a field, similar to $B$-boson field.

In Part 4 {\it operators of a particles creation} and {\it a particles
annihilation} are denoted as the Fourier transformations of the
corresponding operators for probability and an antiparticles are denoted in
the standard way.

In Part 5 here are considered all unitary transformations on the
two-masses functions such that these transformations retain the probability
4-vector. The adequate to electroweak gauge fields transformations exist
among these unitary transformations. These electroweak unitary
transformations are expressed by rotations at the 2-space of the fourth and
the fifth coordinates, too.

The motion equations are invariant for these transformations. The massless
field $W_{\mu ,\nu }$ is denoted in usual way, but the motion equations for
the fields $W_\mu $ are similar to the Klein-Gordon equation with nonzero
mass.

The massless field $A$ and the massive field $Z$ are denoted in standard way
by the fields $B$ and $W$.

\section{Hamiltonians}

Let us denote:

$\mathbf{e}_1$, $\mathbf{e}_2$, $\mathbf{e}_3$ are the Cartesian basis
vectors;

$\mathbf{x}\ \stackrel{def}{=}\left( x_1\mathbf{e}_1+x_2\mathbf{e}_2+x_3%
\mathbf{e}_3\right) $;

$x_0\ \stackrel{def}{=}t$;

$\int d^3\mathbf{x}\ \stackrel{def}{=}\int dx_1\int dx_2\int dx_3$;

$\partial _k\ \stackrel{def}{=}\partial /\partial x_k$;

$\partial _t\ \stackrel{def}{=}\partial _0\ \stackrel{def}{=}\partial
/\partial t$;

$\partial _k^{\prime }\ \stackrel{def}{=}\partial /\partial x_k^{\prime }$.

Let $\wp \left( t,\mathbf{x}\right) $ be any occurred in point $\left( t,%
\mathbf{x}\right) $ event and let a real function $\rho _\wp \left( t,%
\mathbf{x}\right) $ be {\it the probability density} of this event. That
is for each domain $D$ ($D\subseteq R^3$):

\[
\int_Dd^3\mathbf{x}\cdot \rho _\wp \left( t,\mathbf{x}\right) =\mathbf{P}%
\left( \exists \mathbf{x}\in D:\wp \left( t,\mathbf{x}\right) \right) 
\]

with $\mathbf{P}$ as a probability function. $\rho _\wp$ is not invariant for the 
Lorentz transformation. Let $\left\langle\rho _\wp,\mathbf{j_\wp}\right\rangle $ be 
a probability current 3+1-vector.

Let  $k\in \left\{ 1,2,3,4\right\} $, $s\in \left\{ 1,2,3,4\right\} $, $%
\alpha \in \left\{ 1,2,3\right\} $ and  $\varphi _k\left( t,\mathbf{x}%
\right) $ be some complex solution of the following set of equations:

\begin{equation}
\left\{ 
\begin{array}{c}
\sum_{k=1}^4\varphi _k^{*}\left( t,\mathbf{x}\right) \varphi _k\left( t,%
\mathbf{x}\right) =\rho _{\wp ,\alpha }\left( t,\mathbf{x}\right) \mbox{,}
\\ 
\sum_{k=1}^4\sum_{s=1}^4\varphi _k^{*}\left( t,\mathbf{x}\right) \beta
_{k,s}^{\left[ \alpha \right] }\varphi _s\left( t,\mathbf{x}\right) =-j_{\wp
,\alpha }\left( t,\mathbf{x}\right) \mbox{.}
\end{array}
\right|   \label{ro}
\end{equation}

If a 3-vector $\mathbf{u}_{\wp}$ is denoted as

\begin{equation}
\mathbf{j}_{\wp}\stackrel{def}{=}\rho _{\wp}\mathbf{u}_{\wp}  \label{v2}
\end{equation}

then  $\mathbf{u}_{\wp}$ is called as {\it a local velocity of the probability
propagation}.

I consider only events, fulfilled to the following condition: there
exists some tiny real positive number $h$ such that if $\left| x_r\right|
\geq \frac \pi h$ ($r\in \left\{ 1,2,3\right\} $) then

\[
\varphi _j\left( t,\mathbf{x}\right) =0\mbox{.} 
\]

Here is suitable to choose this number such that $h$ equals the Planck
constant for our devices. But maybe somewhere exist devices of other
structure such that another value for $h$ will required.

Let $\left( V\right) $ be denoted as the following: $\mathbf{x}\in \left(
V\right) $ if and only if $\left| x_r\right| \leq \frac \pi h$ for $r\in
\left\{ 1,2,3\right\} $. That is:

\[
\int_{\left( V\right) }d^3\mathbf{x}=\int_{-\frac \pi h}^{\frac \pi
h}dx_1\int_{-\frac \pi h}^{\frac \pi h}dx_2\int_{-\frac \pi h}^{\frac \pi
h}dx_3\mbox{.} 
\]

Let $j\in \left\{ 1,2,3,4\right\} $, $k\in \left\{ 1,2,3,4\right\} $

Let

\[
\varphi _j\left( t,\mathbf{x}\right) =\sum_{w,\mathbf{p}}c_{j,w,\mathbf{p}%
}\varsigma _{w,\mathbf{p}}\left( t,\mathbf{x}\right) 
\]

with $\varsigma _{w,\mathbf{p}}\left( t,\mathbf{x}\right) \stackrel{def}{=}%
\exp \left( \mathrm{i}h\left( wt-\mathbf{px}\right) \right) $ be the Fourier
series for $\varphi _j\left( t,\mathbf{x}\right) $.

Let $\varphi _{j,w,\mathbf{p}}\left( t,\mathbf{x}\right) \stackrel{def}{=}%
c_{j,w,\mathbf{p}}\varsigma _{w,\mathbf{p}}\left( t,\mathbf{x}\right) $.

Let $\left\langle t,\mathbf{x}\right\rangle $ be any space-time point.

Denote value of function $\varphi _{k,w,\mathbf{p}}$ at this point as

\[
\varphi _{k,w,\mathbf{p}}|_{\left\langle t,\mathbf{x}\right\rangle }=A_k
\]

and value of function $\partial _t\varphi _{j,w,\mathbf{p}%
}-\sum_{s=1}^4\sum_{\alpha =1}^3\beta _{j,s}^{\left[ \alpha \right]
}\partial _\alpha \varphi _{s,w,\mathbf{p}}$ at this point as 

\[
\left( \partial _t\varphi _{j,w,\mathbf{p}}-\sum_{s=1}^4\sum_{\alpha
=1}^3\beta _{j,s}^{\left[ \alpha \right] }\partial _\alpha \varphi _{s,w,%
\mathbf{p}}\right) |_{\left\langle t,\mathbf{x}\right\rangle }=C_j.
\]

There $A_k$ and $C_j$ are complex numbers. Hence the following equations set:

\begin{equation}
\left\{ 
\begin{array}{c}
\sum_{k=1}^4z_{j,k,w,\mathbf{p}}A_k=C_j\mbox{,} \\ 
z_{j,k,w,\mathbf{p}}^{*}=-z_{k,j,w,\mathbf{p}}
\end{array}
\right|  \label{sys}
\end{equation}

is a set of 20 algebraic complex equations with 16 complex unknown numbers $%
z_{k,j,w,\mathbf{p}}$. This set can be reformulated as the set of 8 linear
real equations with 16 real unknown numbers $\mathrm{Re}\left( z_{j,k,w,%
\mathbf{p}}\right) $ for $j<k$ and $\mathrm{Im}\left( z_{j,k,w,\mathbf{p}%
}\right) $ for $j\leq k$. This set has got solutions by the
Kronecker-Capelli theorem. Hence at every point $\left\langle t,\mathbf{x}%
\right\rangle $ such complex number $z_{j,k,w,\mathbf{p}}|_{\left\langle t,%
\mathbf{x}\right\rangle }$ exists.

Let $\kappa _{w,\mathbf{p}}$ be a linear operator on the linear space,
spanned by functions $\varsigma _{w,\mathbf{p}}\left( t,\mathbf{x}\right) $,
and

\[
\kappa _{w,\mathbf{p}}\varsigma _{w^{\prime },\mathbf{p}^{\prime }}\stackrel{%
def}{=}\left\{ 
\begin{array}{c}
\varsigma _{w^{\prime },\mathbf{p}^{\prime }}\mbox{, if }w=w^{\prime }%
\mbox{, }\mathbf{p}=\mathbf{p}^{\prime }\mbox{;} \\ 
0\mbox{, if }w\neq w^{\prime }\mbox{or/and }\mathbf{p}\neq \mathbf{p}%
^{\prime }
\end{array}
\mbox{.}\right| 
\]

Let $Q_{j,k}$ be a operator such that in every point $\left\langle t,\mathbf{%
x}\right\rangle $:

\[
Q_{j,k}|_{\left\langle t,\mathbf{x}\right\rangle }\stackrel{def}{=}\sum_{w,%
\mathbf{p}}\left( z_{j,k,w,\mathbf{p}}|_{\left\langle t,\mathbf{x}%
\right\rangle }\right) \kappa _{w,\mathbf{p}} 
\]

Therefore for every $\varphi $ there exist operator $Q_{j,k}$ such that the $%
\varphi $ dependence upon $t$ is characterized by the following differential
equations\footnote{%
This set of equations is similar to the Dirac equation with the mass matrix 
\cite{VVD}, \cite{Barut}, \cite{Wilson}. I choose form of this set of
equations in order to describe behaviour of $\rho _\wp \left( t,\mathbf{x}%
\right) $ by spinors and by the Clifford's set elements.}:

\begin{equation}
\partial _t\varphi _j=\sum_{k=1}^4\left( \beta _{j,k}^{\left[ 1\right]
}\partial _1+\beta _{j,k}^{\left[ 2\right] }\partial _2+\beta _{j,k}^{\left[
3\right] }\partial _3+Q_{j,k}\right) \varphi _k\mbox{.}  \label{ham}
\end{equation}

and $Q_{j,k}^{*}=\sum_{w,\mathbf{p}}\left( z_{j,k,w,\mathbf{p}%
}^{*}|_{\left\langle t,\mathbf{x}\right\rangle }\right) \kappa _{w,\mathbf{p}%
}=-Q_{k,j}$.

In that case if

\[
\widehat{H}_{j,k}\stackrel{def}{=}\mathrm{i}\left( \beta _{j,k}^{\left[
1\right] }\partial _1+\beta _{j,k}^{\left[ 2\right] }\partial _2+\beta
_{j,k}^{\left[ 3\right] }\partial _3+Q_{j,k}\right) 
\]

then $\widehat{H}$ is called as {\it the hamiltonian} of the moving with
equation (\ref{ham}).

Let $\mathbf{H}$ be some Hilbert space such that some linear operators $\psi
_j\left( \mathbf{x}\right) $ act on elements of $\mathbf{H}$. And these
operators have got the following properties:

1. $\mathbf{H}$ contains the element $\Phi _0$ such that:

\[
\Phi _0^{\dagger }\Phi _0=1 
\]

and

\[
\psi _j\Phi _0=0\mbox{, }\Phi _0^{\dagger }\psi _j^{\dagger }=0\mbox{;} 
\]

2.

\[
\psi _j\left( \mathbf{x}\right) \psi _j\left( \mathbf{x}\right) =0 
\]

and

\[
\psi _j^{\dagger }\left( \mathbf{x}\right) \psi _j^{\dagger }\left( \mathbf{x%
}\right) =0\mbox{;} 
\]

3.

\begin{equation}
\begin{array}{c}
\left\{ \psi _{j^{\prime }}^{\dagger }\left( \mathbf{y}\right) ,\psi
_j\left( \mathbf{x}\right) \right\} \stackrel{Def}{=} \\ 
\stackrel{Def}{=}\psi _{j^{\prime }}^{\dagger }\left( \mathbf{y}\right) \psi
_j\left( \mathbf{x}\right) +\psi _j\left( \mathbf{x}\right) \psi _{j^{\prime
}}^{\dagger }\left( \mathbf{y}\right) =\delta \left( \mathbf{y}-\mathbf{x}%
\right) \delta _{j^{\prime },j}
\end{array}
\label{dddd}
\end{equation}

Let us denote $\Psi \left( t,\mathbf{x}\right) $ as the following:

\begin{equation}
\Psi \left( t,\mathbf{x}\right) \stackrel{def}{=}\sum_{j=1}^4\varphi
_j\left( t,\mathbf{x}\right) \psi _j^{\dagger }\left( \mathbf{x}\right) \Phi
_0  \label{Sat}
\end{equation}

From (\ref{dddd}):

\[
\Psi ^{\dagger }\left( t,\mathbf{x}^{\prime }\right) \Psi \left( t,\mathbf{x}%
\right) =\sum_{j=1}^4\varphi _j^{*}\left( t,\mathbf{x}^{\prime }\right)
\varphi _j\left( t,\mathbf{x}\right) \delta \left( \mathbf{x}^{\prime }-%
\mathbf{x}\right) \mbox{.} 
\]

That is from (\ref{ro}):

\[
\int dx^{\prime }\cdot \Psi ^{\dagger }\left( t,\mathbf{x}^{\prime }\right)
\Psi \left( t,\mathbf{x}\right) =\rho _\wp \left( t,\mathbf{x}\right) \mbox{.}
\]

I call operator $\psi ^{\dagger }\left( \mathbf{x}\right) $ as{\it \ a
creation operator} and $\psi \left( \mathbf{x}\right) $ is called as {\it %
an annihilation operator of the event }$\wp $ {\it probability at point} $%
\mathbf{x}$. Operator $\psi ^{\dagger }\left( \mathbf{x}\right) $ is not an 
operator of a particle creation at point $\mathbf{x}$ but it is an operator, 
changing probability of event $\wp $ at this point. The similar is for 
$\psi \left( \mathbf{x}\right) $.

If $\mathcal{H}\left( t,\mathbf{x}\right) $ is denoted as:

\begin{equation}
\mathcal{H}\left( t,\mathbf{x}\right) \stackrel{def}{=}\sum_{j=1}^4\psi
_j^{\dagger }\left( \mathbf{x}\right) \sum_{k=1}^4\widehat{H}_{j,k}\left( t,%
\mathbf{x}\right) \psi _k\left( \mathbf{x}\right)  \label{hmm}
\end{equation}

then $\mathcal{H}\left(t,\mathbf{x}\right) $ is called as {\it the
hamiltonian density} of this hamiltonian.

From (\ref{Sat}):

\[
-\mathrm{i}\int d^3\mathbf{x}\cdot \mathcal{H}\left( t,\mathbf{x}\right)
\Psi \left( t,\mathbf{x}_0\right) =\partial _t\Psi \left( t,\mathbf{x}%
_0\right) \mbox{.} 
\]

Therefore a hamiltonian density defines the temporal behaviour of probability at a 
space point.

Formula (\ref{ham}) has got the following matrix form:

\begin{equation}
\partial _t\varphi =\left( \beta ^{\left[ 1\right] }\partial _1+\beta
^{\left[ 2\right] }\partial _2+\beta ^{\left[ 3\right] }\partial _3+\widehat{%
Q}\right) \varphi \mbox{,}  \label{ham1}
\end{equation}

with

\[
\varphi =\left[ 
\begin{array}{c}
\varphi _1 \\ 
\varphi _2 \\ 
\varphi _3 \\ 
\varphi _4
\end{array}
\right] 
\]

and

\[
\widehat{Q}=\left[ 
\begin{array}{cccc}
\mathrm{i}\vartheta _{1,1} & \mathrm{i}\vartheta _{1,2}-\varpi _{1,2} & 
\mathrm{i}\vartheta _{1,3}-\varpi _{1,3} & \mathrm{i}\vartheta _{1,4}-\varpi
_{1,4} \\ 
\mathrm{i}\vartheta _{1,2}+\varpi _{1,2} & \mathrm{i}\vartheta _{2,2} & 
\mathrm{i}\vartheta _{2,3}-\varpi _{2,3} & \mathrm{i}\vartheta _{2,4}-\varpi
_{2,4} \\ 
\mathrm{i}\vartheta _{1,3}+\varpi _{1,3} & \mathrm{i}\vartheta _{2,3}+\varpi
_{2,3} & \mathrm{i}\vartheta _{3,3} & \mathrm{i}\vartheta _{3,4}-\varpi
_{3,4} \\ 
\mathrm{i}\vartheta _{1,4}+\varpi _{1,4} & \mathrm{i}\vartheta _{2,4}+\varpi
_{2,4} & \mathrm{i}\vartheta _{3,4}+\varpi _{3,4} & \mathrm{i}\vartheta
_{4,4}
\end{array}
\right] 
\]

with $\varpi _{j,k}={\rm Re}\left( Q_{j,k}\right) $ and $\vartheta _{1,2}=%
{\rm Im}\left(Q_{j,k}\right) $.

Let $\Theta _0$, $\Theta _3$, $\Upsilon _0$ and $\Upsilon _3$ be the
solution of the following system of equations:

\[
\left\{ 
\begin{array}{c}
{-\Theta _0+\Theta _3-\Upsilon _0+\Upsilon _3}{}{=\vartheta _{1,1}}\mbox{;}
\\ 
{-\Theta _0-\Theta _3-\Upsilon _0-\Upsilon _3}{}{=\vartheta _{2,2}}\mbox{;}
\\ 
{-\Theta _0-\Theta _3+\Upsilon _0+\Upsilon _3}{}{=\vartheta _{3,3}}\mbox{;}
\\ 
{-\Theta _0+\Theta _3+\Upsilon _0-\Upsilon _3}{}{=\vartheta _{4,4}}
\end{array}
\right| \mbox{.} 
\]

Let $\Theta _1$, $\Upsilon _1$, $\Theta _2$, $\Upsilon _2$, ${M_0}$, ${M_4}$%
, ${M_{1,0}}$, ${M_{1,4}}$, ${M_{2,0}}$, ${M_{2,4}}$, ${M_{3,0}}$, ${M_{3,4}}
$ be the solutions of the following systems of equations:

\[
\left\{ 
\begin{array}{c}
{\ \Theta _1+\Upsilon _1}{}{=\vartheta _{1,2}}\mbox{;} \\ 
{-\Theta _1+\Upsilon _1}{}{=\vartheta _{3,4}}\mbox{;}
\end{array}
\right| 
\]

\[
\left\{ 
\begin{array}{c}
{-\Theta _2-\Upsilon _2}{}{=\varpi _{1,2}}\mbox{;} \\ 
{\Theta _2-\Upsilon _2}{}{=\varpi _{3,4}}\mbox{;}
\end{array}
\right| 
\]

\[
\left\{ 
\begin{array}{c}
{M_0+M_{3,0}}{}{=\vartheta _{1,3}}\mbox{;} \\ 
{M_0-M_{3,0}}{}{=\vartheta _{2,4}}\mbox{;}
\end{array}
\right| 
\]

\[
\left\{ 
\begin{array}{c}
{M_4+M_{3,4}}{}{=\varpi _{1,3}}\mbox{;} \\ 
{M_4-M_{3,4}}{}{=\varpi _{2,4}}\mbox{;}
\end{array}
\right| 
\]

\[
\left\{ 
\begin{array}{c}
{M_{1,0}-M_{2,4}}{}{=\vartheta _{1,4}}\mbox{;} \\ 
{M_{1,0}+M_{2,4}}{}{=\vartheta _{2,3}}\mbox{;}
\end{array}
\right| 
\]

\[
\left\{ 
\begin{array}{c}
{M_{1,4}-M_{2,0}}{}{=\varpi _{1,4}}\mbox{;} \\ 
{M_{1,4}+M_{2,0}}{}{=\varpi _{2,3}}
\end{array}
\right|\mbox{.} 
\]

From (\ref{ham1}):

\begin{eqnarray}
&&\left( \partial _t+\mathrm{i}\Theta _0+\mathrm{i}\Upsilon _0\gamma
^{\left[ 5\right] }\right) \varphi =  \nonumber \\
&=&\left( 
\begin{array}{c}
\sum_{k=1}^3\beta ^{\left[ k\right] }\left( \partial _k+\mathrm{i}\Theta _k+%
\mathrm{i}\Upsilon _k\gamma ^{\left[ 5\right] }\right) +\mathrm{i}M_0\gamma
^{\left[ 0\right] }+\mathrm{i}M_4\beta ^{\left[ 4\right] } \\ 
-\mathrm{i}M_{1,0}\gamma _\zeta ^{\left[ 0\right] }-\mathrm{i}M_{1,4}\zeta
^{\left[ 4\right] }- \\ 
-\mathrm{i}M_{2,0}\gamma _\eta ^{\left[ 0\right] }-\mathrm{i}M_{2,4}\eta
^{\left[ 4\right] }- \\ 
-\mathrm{i}M_{3,0}\gamma _\theta ^{\left[ 0\right] }-\mathrm{i}M_{3,4}\theta
^{\left[ 4\right] }
\end{array}
\right) \varphi \label{ham0}
\end{eqnarray}

with

\[
\gamma ^{\left[ 5\right] }\stackrel{def}{=}\left[ 
\begin{array}{cc}
1_2 & 0_2 \\ 
0_2 & -1_2
\end{array}
\right] \mbox{.}
\]

Here summands

\[
\begin{array}{c}
-\mathrm{i}M_{1,0}\gamma _\zeta ^{\left[ 0\right] }-\mathrm{i}M_{1,4}\zeta
^{\left[ 4\right] }- \\ 
-\mathrm{i}M_{2,0}\gamma _\eta ^{\left[ 0\right] }-\mathrm{i}M_{2,4}\eta
^{\left[ 4\right] }- \\ 
-\mathrm{i}M_{3,0}\gamma _\theta ^{\left[ 0\right] }-\mathrm{i}M_{3,4}\theta
^{\left[ 4\right] }
\end{array}
\]

contain the chromatic pentads elements \cite{Q42} and

\[
\sum_{k=1}^3\beta ^{\left[ k\right] }\left( \partial _k+\mathrm{i}\Theta _k+%
\mathrm{i}\Upsilon _k\gamma ^{\left[ 5\right] }\right) +\mathrm{i}M_0\gamma
^{\left[ 0\right] }+\mathrm{i}M_4\beta ^{\left[ 4\right] } 
\]

contains only the light pentad elements. I call the following sum

\begin{equation}
\widehat{H}_l\stackrel{def}{=}\sum_{k=1}^3\beta ^{\left[ k\right] }\left( 
\mathrm{i}\partial _k-\Theta _k-\Upsilon _k\gamma ^{\left[ 5\right] }\right)
-M_0\gamma ^{\left[ 0\right] }-M_4\beta ^{\left[ 4\right] } \label{oh}
\end{equation}

as {\it the leptonn hamiltonian}.

\section{Rotation of $x_5Ox_4$ and $B$-bosonn}

If denote (\ref{ro}):

\[
j_4\stackrel{def}{=}-\varphi ^{*}\beta ^{\left[ 4\right] }\varphi 
\mbox{ and
}j_5\stackrel{def}{=}-\varphi ^{*}\gamma ^{\left[ 0\right] }\varphi 
\]

and (\ref{v2}):

\[
\rho _\wp \left( t,\mathbf{x}\right)u_4\stackrel{def}{=}j_4\mbox{ and }\rho
_\wp \left( t,\mathbf{x}\right)u_5\stackrel{def}{=}j_5\mbox{,} 
\]

then

\[
u_1^2+u_2^2+u_3^2+u_4^2+u_5^2=1\mbox{.} 
\]

Hence, only all the five elements of the Clifford pentad lend the entire kit
of the velocity components. Two more ''space'' coordinates $x_5$ and $x_4$
should be added to our three $x_1,x_2,x_3$ for the completeness. These
additional coordinates can be chosen such that

\[
-\frac \pi h\leq x_5\leq \frac \pi h,-\frac \pi h\leq x_4\leq \frac \pi h%
\mbox{.} 
\]

$x_4$ and $x_5$ are not coordinates of any physics events. Hence our devices
do not detect these coordinates as our space coordinates. 

Let us denote:

\begin{eqnarray}
&&\widetilde{\varphi }\left( t,x_1,x_2,x_3,x_5,x_4\right) \stackrel{def}{=}%
\varphi \left( t,x_1,x_2,x_3\right) \cdot  \nonumber \\
&&\cdot \left( \exp \left( -\mathrm{i}\left( x_5M_0\left(
t,x_1,x_2,x_3\right) +x_4M_4\left( t,x_1,x_2,x_3\right) \right) \right)
\right) \mbox{.}  \nonumber
\end{eqnarray}

In this case the motion equation for the leptonn hamiltonian (\ref{oh}) is the
following:

\begin{equation}
\left( \sum_{\mu =0}^3\beta ^{\left[ \mu \right] }\left( \mathrm{i}\partial
_\mu -\Theta _\mu -\Upsilon _\mu \gamma ^{\left[ 5\right] }\right) +\gamma
^{\left[ 0\right] }\mathrm{i}\partial _5+\beta ^{\left[ 4\right] }\mathrm{i}%
\partial _4\right) \widetilde{\varphi }=0.  \label{gkk}
\end{equation}

Let $g_1$ be some positive real number and for $\mu \in \left\{
0,1,2,3\right\} $: $F_\mu $ and $B_\mu $ be the solutions of the following
system of the equations:

\[
\left\{ 
\begin{array}{c}
{-0.5g_1B_\mu +F_\mu }{}{=-\Theta _\mu -\Upsilon _\mu ,}\mbox{;} \\ 
{-g_1B_\mu +F_\mu }{}{=-\Theta _\mu +\Upsilon _\mu }\mbox{.}
\end{array}
\right| 
\]

Let {\it the charge matrix} be denoted as the following:

\[
Y\stackrel{def}{=}-\left[ 
\begin{array}{cc}
1_2 & 0_2 \\ 
0_2 & 2\cdot 1_2
\end{array}
\right] \mbox{.} 
\]

Hence from (\ref{gkk}):

\begin{equation}
\left( \sum_{\mu =0}^3\beta ^{\left[ \mu \right] }\left( \mathrm{i}\partial
_\mu +F_\mu +0.5g_1YB_\mu \right) +\gamma ^{\left[ 0\right] }\mathrm{i}%
\partial _5+\beta ^{\left[ 4\right] }\mathrm{i}\partial _4\right) \widetilde{%
\varphi }=0\mbox{.}  \label{gkB}
\end{equation}

Let $\chi \left( t,x_1,x_2,x_3\right) $ be any real function and:

\begin{equation}
\widetilde{U}\left( \chi \right) \stackrel{def}{=}\left[ 
\begin{array}{cc}
\exp \left( \mathrm{i}\frac \chi 2\right) 1_2 & 0_2 \\ 
0_2 & \exp \left( \mathrm{i}\chi \right) 1_2
\end{array}
\right] \mbox{.}  \label{ux}
\end{equation}

The motion equation (\ref{gkB}) is invariant for the following
transformation (rotation of $x_4Ox_5$):

\begin{eqnarray}
&&x_4\rightarrow x_4^{\prime }=x_4\cos \frac \chi 2-x_5\sin \frac \chi 2%
\mbox{;}  \nonumber \\
&&x_5\rightarrow x_5^{\prime }=x_5\cos \frac \chi 2+x_4\sin \frac \chi 2%
\mbox{;}  \nonumber \\
&&x_\mu \rightarrow x_\mu ^{\prime }=x_\mu \mbox{ for
}\mu \in \left\{ 0,1,2,3\right\} \mbox{;}  \nonumber \\
&&Y\rightarrow Y^{\prime }=\widetilde{U}^{\dagger }Y\widetilde{U}=Y\mbox{;}
\label{T} \\
&&\widetilde{\varphi }\rightarrow \widetilde{\varphi }^{\prime }=\widetilde{U%
}\widetilde{\varphi }\mbox{,}  \nonumber \\
&&B_\mu \rightarrow B_\mu ^{\prime }=B_\mu -\frac 1{g_1}\partial _\mu \chi %
\mbox{,}  \nonumber \\
&&F_\mu \rightarrow F_\mu ^{\prime }=F_\mu \mbox{.}  \nonumber
\end{eqnarray}

Hence $B_\mu $ is similar to the $B$-boson field from Standard Model. I call
it as $B$-bosonn.

Let $\epsilon _\mu $ ($\mu \in \left\{ 1,2,3,4\right\} $) be the basis such
that in this basis the light pentad has got the form (\ref{lghr}) .

Spinor functions of type

\[
\frac h{2\pi }\exp \left( -\mathrm{i}h\left( sx_4+nx_5\right) \right)
\epsilon _k 
\]

with integer $n$ and $s$ make up an orthonormal basis of some space (let us
denote this space as $\Im $) with the following scalar product:

\begin{equation}
\widetilde{\varphi }*\widetilde{\chi }\stackrel{def}{=}\int_{-\frac \pi
h}^{\frac \pi h}dx_5\int_{-\frac \pi h}^{\frac \pi h}dx_4\cdot \left( 
\widetilde{\varphi }^{\dagger }\cdot \widetilde{\chi }\right) \mbox{.}
\label{sp}
\end{equation}

In this case from (\ref{ro}):

\begin{equation}
\widetilde{\varphi }*\beta ^{\left[ \mu \right] }\widetilde{\varphi }=-j_{\wp,\mu}
\label{jax}
\end{equation}

for $\mu \in \left\{ 0,1,2,3\right\} $.

The Fourier series for $\widetilde{\varphi }\left( t,\mathbf{x}%
,x_5,x_4\right) $ has got the following form:

\begin{equation}
\begin{array}{c}
\widetilde{\varphi }\left( t,\mathbf{x},x_5,x_4\right) = \\ 
=\sum_{n,s}\phi \left( t,\mathbf{x},n,s\right) \frac h{2\pi }\exp \left( -%
\mathrm{i}h\left( nx_5+sx_4\right) \right) \mbox{.}
\end{array}
\label{lt}
\end{equation}

The integer numbers $n$ and $s$ be called as{\it \ mass numbers} for $%
\widetilde{\varphi }$ and $\sqrt{n^2+s^2}$ is called as the mass for $%
\widetilde{\varphi }$.

\section{The one-mass state, particles and antiparticles.}

Let (\ref{lt}):

\[
\widetilde{\varphi }\left( t,\mathbf{x},x_5,x_4\right) =\exp \left( -\mathrm{%
i}hnx_5\right) \sum_{j=1}^4\phi _j\left( t,\mathbf{x},n,0\right) \epsilon _j 
\]

with $n$ as some natural number. In that case the hamiltonian has got the
following form from (\ref{gkB}):

\[
\widehat{H}=\sum_{k=1}^3\beta ^{\left[ k\right] }\mathrm{i}\partial
_k+hn\gamma ^{\left[ 0\right] }+\widehat{G} 
\]

with

\[
\widehat{G}\stackrel{def}{=}\sum_{\mu =0}^3\beta ^{\left[ \mu \right]
}\left( F_\mu +0.5g_1YB_\mu \right) \mbox{.} 
\]

If

\begin{equation}
\widehat{H}_0\stackrel{def}{=}\sum_{k=1}^3\beta ^{\left[ k\right] }\mathrm{i}%
\partial _k+hn\gamma ^{\left[ 0\right] }  \label{hmm0}
\end{equation}

then the functions

\[
u_1\left( \mathbf{k}\right) \exp \left( -\mathrm{i}h\mathbf{kx}\right) 
\mbox{
and }u_2\left( \mathbf{k}\right) \exp \left( -\mathrm{i}h\mathbf{kx}\right) 
\]

with

\[
u_1\left( \mathbf{k}\right) \stackrel{def}{=}\frac 1{2\sqrt{\omega \left( 
\mathbf{k}\right) \left( \omega \left( \mathbf{k}\right) +n\right) }}\left[ 
\begin{array}{c}
\omega \left( \mathbf{k}\right) +n+k_3 \\ 
k_1+\mathrm{i}k_2 \\ 
\omega \left( \mathbf{k}\right) +n-k_3 \\ 
-k_1-\mathrm{i}k_2
\end{array}
\right] 
\]

and

\[
u_2\left( \mathbf{k}\right) \stackrel{def}{=}\frac 1{2\sqrt{\omega \left( 
\mathbf{k}\right) \left( \omega \left( \mathbf{k}\right) +n\right) }}\left[ 
\begin{array}{c}
k_1-\mathrm{i}k_2 \\ 
\omega \left( \mathbf{k}\right) +n-k_3 \\ 
-k_1+\mathrm{i}k_2 \\ 
\omega \left( \mathbf{k}\right) +n+k_3
\end{array}
\right] 
\]

are the eigenvectors of $\widehat{H}_0$ with the eigenvalue $\omega \left( 
\mathbf{k}\right) \ \stackrel{def}{=}\sqrt{\mathbf{k}^2+n^2}$ and the
functions

\[
u_3\left( \mathbf{k}\right) \exp \left( -\mathrm{i}h\mathbf{kx}\right) 
\mbox{
and }u_4\left( \mathbf{k}\right) \exp \left( -\mathrm{i}h\mathbf{kx}\right) 
\]

with

\[
u_3\left( \mathbf{k}\right) \stackrel{def}{=}\frac 1{2\sqrt{\omega \left( 
\mathbf{k}\right) \left( \omega \left( \mathbf{k}\right) +n\right) }}\left[ 
\begin{array}{c}
-\omega \left( \mathbf{k}\right) -n+k_3 \\ 
k_1+\mathrm{i}k_2 \\ 
\omega \left( \mathbf{k}\right) +n+k_3 \\ 
k_1+\mathrm{i}k_2
\end{array}
\right] 
\]

and

\[
u_4\left( \mathbf{k}\right) \stackrel{def}{=}\frac 1{2\sqrt{\omega \left( 
\mathbf{k}\right) \left( \omega \left( \mathbf{k}\right) +n\right) }}\left[ 
\begin{array}{c}
k_1-\mathrm{i}k_2 \\ 
-\omega \left( \mathbf{k}\right) -n-k_3 \\ 
k_1-\mathrm{i}k_2 \\ 
\omega \left( \mathbf{k}\right) +n-k_3
\end{array}
\right] 
\]

are the eigenvectors of $\widehat{H}_0$ with the eigenvalue $-\omega \left( 
\mathbf{k}\right) $.

Here $u_\mu \left( \mathbf{k}\right) $ form an orthonormal basis in the space,
spanned on vectors $\epsilon _\mu $.

Let :

\[
b_{r,\mathbf{k}}\stackrel{def}{=}\left( \frac h{2\pi }\right)
^3\sum_{j^{\prime }=1}^4\int_{\left( V\right) }d^3\mathbf{x}^{\prime }\cdot
e^{\mathrm{i}h\mathbf{kx}^{\prime }}u_{r,j^{\prime }}^{*}\left( \mathbf{k}%
\right) \psi _{j^{\prime }}\left( \mathbf{x}^{\prime }\right) 
\]

In that case

\begin{equation}
\psi _j\left( \mathbf{x}\right) =\sum_{\mathbf{k}}e^{-\mathrm{i}h\mathbf{kx}%
}\sum_{r=1}^4b_{r,\mathbf{k}}u_{r,j}\left( \mathbf{k}\right)  \label{bb}
\end{equation}

with

\[
\sum_{\mathbf{k}}\stackrel{def}{=}\sum_{k_1=-\infty }^\infty
\sum_{k_2=-\infty }^\infty \sum_{k_3=-\infty }^\infty \mbox{;} 
\]

and in this case

\begin{eqnarray}
&&\left\{ b_{s,\mathbf{k}^{\prime }}^{\dagger },b_{r,\mathbf{k}}\right\}
=\left( \frac h{2\pi }\right) ^3\delta _{s,r}\delta _{\mathbf{k},\mathbf{k}%
^{\prime }}\mbox{,}  \nonumber \\
&&\left\{ b_{s,\mathbf{k}^{\prime }}^{\dagger },b_{r,\mathbf{k}}^{\dagger
}\right\} =0=\left\{ b_{s,\mathbf{k}^{\prime }},b_{r,\mathbf{k}}\right\} %
\mbox{,}  \label{bubu} \\
&&b_{r,\mathbf{k}}\Phi _0=0\mbox{.}  \nonumber
\end{eqnarray}

The hamiltonian density (\ref{hmm}) for $\widehat{H}_0$ is the following:

\[
\mathcal{H}_0\left( \mathbf{x}\right) =\sum_{j=1}^4\psi _j^{\dagger }\left( 
\mathbf{x}\right) \sum_{k=1}^4\widehat{H}_{0,j,k}\psi _k\left( \mathbf{x}%
\right) \mbox{.} 
\]

Hence from (\ref{bb}):

\[
\int_{\left( V\right) }d^3\mathbf{x}\cdot \mathcal{H}_0\left( \mathbf{x}%
\right) =\left( \frac{2\pi }h\right) ^3\sum_{\mathbf{k}}h\omega \left( 
\mathbf{k}\right) \cdot \left( \sum_{r=1}^2b_{r,\mathbf{k}}^{\dagger }b_{r,%
\mathbf{k}}-\sum_{r=3}^4b_{r,\mathbf{k}}^{\dagger }b_{r,\mathbf{k}}\right) 
\]

Let the Fourier transformation for $\varphi $ be the following:

\[
\varphi _j\left( t,\mathbf{x}\right) =\sum_{\mathbf{p}}\sum_{r=1}^4c_r\left(
t,\mathbf{p}\right) u_{r,j}\left( \mathbf{p}\right) e^{-\mathrm{i}h\mathbf{px%
}} 
\]

with

\[
c_r\left( t,\mathbf{p}\right) \stackrel{def}{=}\left( \frac h{2\pi }\right)
^3\sum_{j^{\prime }=1}^4\int_{\left( V\right) }d^3\mathbf{x}^{\prime }\cdot
u_{r,j^{\prime }}^{*}\left( \mathbf{p}\right) \varphi _{j^{\prime }}\left( t,%
\mathbf{x}^{\prime }\right) e^{\mathrm{i}h\mathbf{px}^{\prime }} 
\]

I call a function $\varphi _j\left( t,\mathbf{x}\right) $ as {\it an
ordinary function} if there exists some real positive number $L$ such that

if $\left| p_1\right| >L$ or/and $\left| p_2\right| >L$ or/and $\left|
p_3\right| >L$ then $c_r\left( t,\mathbf{p}\right) =0$.

In that case I denote:

\[
\sum_{\mathbf{p\in \Xi }}\stackrel{def}{=}\sum_{p_1=-L}^L\sum_{p_2=-L}^L%
\sum_{p_3=-L}^L\mbox{.} 
\]

If $\varphi _j\left( t,\mathbf{x}\right) $ is an ordinary function then:

\[
\varphi _j\left( t,\mathbf{x}\right) =\sum_{\mathbf{p\in \Xi }%
}\sum_{r=1}^4c_r\left( t,\mathbf{p}\right) u_{r,j}\left( \mathbf{p}\right)
e^{-\mathrm{i}h\mathbf{px}}\mbox{.} 
\]

Hence from (\ref{Sat}):

\[
\Psi \left( t,\mathbf{x}\right) =\sum_{\mathbf{p}}\sum_{r=1}^4\sum_{\mathbf{k%
}}\sum_{r^{\prime }=1}^4c_r\left( t,\mathbf{p}\right) e^{\mathrm{i}h\left( 
\mathbf{k}-\mathbf{p}\right) \mathbf{x}}\sum_{j=1}^4u_{r^{\prime
},j}^{*}\left( \mathbf{k}\right) u_{r,j}\left( \mathbf{p}\right)
b_{r^{\prime },\mathbf{k}}^{\dagger }\Phi _0 
\]

and

\[
\int_{\left( V\right) }d^3\mathbf{x}\cdot \Psi \left( t,\mathbf{x}\right)
=\left( \frac{2\pi }h\right) ^3\sum_{\mathbf{p}}\sum_{r=1}^4c_r\left( t,%
\mathbf{p}\right) b_{r,\mathbf{p}}^{\dagger }\Phi _0\mbox{.} 
\]

If denote:

\[
\widetilde{\Psi }\left( t,\mathbf{p}\right) \stackrel{def}{=}\left( \frac{%
2\pi }h\right) ^3\sum_{r=1}^4c_r\left( t,\mathbf{p}\right) b_{r,\mathbf{p}%
}^{\dagger }\Phi _0 
\]

then

\[
\int_{\left( V\right) }d^3\mathbf{x}\cdot \Psi \left( t,\mathbf{x}\right)
=\sum_{\mathbf{p}}\widetilde{\Psi }\left( t,\mathbf{p}\right) 
\]

and

\[
H_0\widetilde{\Psi }\left( t,\mathbf{p}\right) =\left( \frac{2\pi }h\right)
^3\sum_{\mathbf{k}}h\omega \left( \mathbf{k}\right) \cdot \left( 
\begin{array}{c}
\sum_{r=1}^2c_r\left( t,\mathbf{k}\right) b_{r,\mathbf{k}}^{\dagger }\Phi _0-
\\ 
-\sum_{r=3}^4c_r\left( t,\mathbf{k}\right) b_{r,\mathbf{k}}^{\dagger }\Phi _0
\end{array}
\right) \mbox{.}
\]

$H_0$ is equivalent to the following operator:

\[
\stackrel{\Xi }{H}_0\stackrel{def}{=}\left( \frac{2\pi }h\right) ^3\sum_{%
\mathbf{k\in \Xi }}h\omega \left( \mathbf{k}\right) \cdot \left(
\sum_{r=1}^2b_{r,\mathbf{k}}^{\dagger }b_{r,\mathbf{k}}-\sum_{r=3}^4b_{r,%
\mathbf{k}}^{\dagger }b_{r,\mathbf{k}}\right) \mbox{.} 
\]

on the set of ordinary functions.

Because from (\ref{bubu})

\[
b_{r,\mathbf{k}}^{\dagger }b_{r,\mathbf{k}}=\left( \frac h{2\pi }\right)
^3-b_{r,\mathbf{k}}b_{r,\mathbf{k}}^{\dagger } 
\]

then

\begin{eqnarray}
\stackrel{\Xi }{H}_0 &=&\left( \frac{2\pi }h\right) ^3\sum_{\mathbf{k\in \Xi 
}}h\omega \left( \mathbf{k}\right) \left( \sum_{r=1}^2b_{r,\mathbf{k}%
}^{\dagger }b_{r,\mathbf{k}}+\sum_{r=3}^4b_{r,\mathbf{k}}b_{r,\mathbf{k}%
}^{\dagger }\right) -  \label{uW} \\
&&-h\sum_{\mathbf{k\in \Xi }}\omega \left( \mathbf{k}\right)\nonumber \mbox{.}
\end{eqnarray}

Let:

\begin{eqnarray}
&&v_{\left( 1\right) }\left( \mathbf{k}\right) \stackrel{def}{=}\gamma
^{\left[ 0\right] }u_3\left( \mathbf{k}\right) \mbox{,}  \nonumber \\
&&v_{\left( 2\right) }\left( \mathbf{k}\right) \stackrel{def}{=}\gamma
^{\left[ 0\right] }u_4\left( \mathbf{k}\right) \mbox{,}  \label{xa} \\
&&u_{\left( 1\right) }\left( \mathbf{k}\right) \stackrel{def}{=}u_1\left( 
\mathbf{k}\right) \mbox{,}  \nonumber \\
&&u_{\left( 2\right) }\left( \mathbf{k}\right) \stackrel{def}{=}u_2\left( 
\mathbf{k}\right)  \nonumber
\end{eqnarray}

and let:

\[
\begin{array}{c}
d_1\left( \mathbf{k}\right) \stackrel{def}{=}-b_3^{\dagger }\left( -\mathbf{k%
}\right) \mbox{,} \\ 
d_2\left( \mathbf{k}\right) \stackrel{def}{=}-b_4^{\dagger }\left( -\mathbf{k%
}\right) \mbox{.}
\end{array}
\]

In that case:

\[
\psi _j\left( \mathbf{x}\right) =\sum_{\mathbf{k}}\sum_{\alpha =1}^2\left(
e^{-\mathrm{i}h\mathbf{kx}}b_{\alpha ,\mathbf{k}}u_{\left( \alpha \right)
,j}\left( \mathbf{k}\right) +e^{\mathrm{i}h\mathbf{kx}}d_{\alpha ,\mathbf{k}%
}^{\dagger }v_{\left( \alpha \right) ,j}\left( \mathbf{k}\right) \right) 
\]

and from (\ref{uW}) the Wick-ordered hamiltonian has got the following form:

\[
:\stackrel{\Xi }{H}_0:=\left( \frac{2\pi }h\right) ^3h\sum_{\mathbf{k\in \Xi 
}}\omega \left( \mathbf{k}\right) \sum_{\alpha =1}^2\left( b_{\alpha ,%
\mathbf{k}}^{\dagger }b_{\alpha ,\mathbf{k}}+d_{\alpha ,\mathbf{k}}^{\dagger
}d_{\alpha ,\mathbf{k}}\right) \mbox{.} 
\]

Here $b_{\alpha ,\mathbf{k}}^{\dagger }$ are called as {\it creation
operators} and $b_{\alpha ,\mathbf{k}}$ are called as {\it annihilation
operators} of $n$-{\it leptonn} with the {\it momentum} $\mathbf{k}$
and the {\it spin index} $\alpha $; $d_{\alpha ,\mathbf{k}}^{\dagger }$
are called as {\it creation operators} and $d_{\alpha ,\mathbf{k}}$ are 
{\it annihilation operators} of {\it anti}-$n$-{\it leptonn} with
the {\it momentum} $\mathbf{k}$ and {\it spin index} $\alpha $.

Functions:

\[
u_{\left( 1\right) }\left( \mathbf{k}\right) \exp \left( -\mathrm{i}h\mathbf{%
kx}\right) \mbox{
and }u_{\left( 2\right) }\left( \mathbf{k}\right) \exp \left( -\mathrm{i}h%
\mathbf{kx}\right) 
\]

are called as {\it the basic} $n$-{\it leptonn functions} with
momentum $\mathbf{k}$ and

\[
v_{\left( 1\right) }\left( \mathbf{k}\right) \exp \left( \mathrm{i}h\mathbf{%
kx}\right) \mbox{
and }v_{\left( 2\right) }\left( \mathbf{k}\right) \exp \left( \mathrm{i}h%
\mathbf{kx}\right) 
\]

are called as {\it the anti-}$n${\it -leptonn basic functions} with
momentum $\mathbf{k}$.

Therefore there arise particles and antiparticles from probabilities of point 
events.

\section{The bi-mass state \cite{DVB}, \cite{AV}}

We consider subspace of space $\Im $ spanned by the following subbasis:

\[
\jmath =\left\langle 
\begin{array}{c}
\frac h{2\pi }\exp \left( -\mathrm{i}h\left( s_0x_4\right) \right) \epsilon
_1,\frac h{2\pi }\exp \left( -\mathrm{i}h\left( s_0x_4\right) \right)
\epsilon _2, \\ 
\frac h{2\pi }\exp \left( -\mathrm{i}h\left( s_0x_4\right) \right) \epsilon
_3,\frac h{2\pi }\exp \left( -\mathrm{i}h\left( s_0x_4\right) \right)
\epsilon _4, \\ 
\frac h{2\pi }\exp \left( -\mathrm{i}h\left( n_0x_5\right) \right) \epsilon
_1,\frac h{2\pi }\exp \left( -\mathrm{i}h\left( n_0x_5\right) \right)
\epsilon _2, \\ 
\frac h{2\pi }\exp \left( -\mathrm{i}h\left( n_0x_5\right) \right) \epsilon
_3,\frac h{2\pi }\exp \left( -\mathrm{i}h\left( n_0x_5\right) \right)
\epsilon _4
\end{array}
\right\rangle 
\]

with some natural $s_0$ and $n_0$. Denote this subspace as $\Im _{\jmath }$.

Let $U$ be any linear transformation at the space $\Im _{\jmath }$ (see
Appendix) such that for any $\widetilde{\varphi }$ : if $\widetilde{\varphi }%
\in \Im _{\jmath }$ then

\begin{equation}
-\left( U\widetilde{\varphi }\right) ^{\dagger }*\beta ^{\left[ \mu \right]
}\left( U\widetilde{\varphi }\right) =j_{\wp,\mu}  \label{uni}
\end{equation}

for $\mu \in \left\{ 0,1,2,3\right\} $ (\ref{jax}).

For every such transformation there exist real functions $\chi \left( t,%
\mathbf{x}\right) $, $\alpha \left( t,\mathbf{x}\right) $, $a\left( t,%
\mathbf{x}\right) $, $b\left( t,\mathbf{x}\right) $, $c\left( t,\mathbf{x}%
\right) $, $q\left( t,\mathbf{x}\right) $, $u\left( t,\mathbf{x}\right) $, $%
v\left( t,\mathbf{x}\right) $, $k\left( t,\mathbf{x}\right) $, $s\left( t,%
\mathbf{x}\right) $ such that

\[
U=\exp \left( \mathrm{i}\alpha \right) \widetilde{U}\left(\chi\right)
U^{\left( -\right) }U^{\left( +\right) } 
\]

with $\widetilde{U}\left( \chi \right) $ as denoted by (\ref{ux}) and $%
U^{\left( -\right) }$ and $U^{\left( +\right) }$ have got the following
matrix form in the basis $\jmath $:

\[
U^{\left( -\right)}\left( t,\mathbf{x}\right)\stackrel = \rm{S}\left( a\left(% 
t,\mathbf{x}\right),b\left( t,\mathbf{x}\right),c\left( t,\mathbf{x}\right),q%
\left( t,\mathbf{x}\right)\right) 
\]

with

\[
a^2\left( t,\mathbf{x}\right)+b^2\left( t,\mathbf{x}\right)+c^2\left( t,\mathbf%
{x}\right)+q^2\left( t,\mathbf{x}\right)=1 
\]

and

\[
\mathrm{S}\left( a,b,c,q\right) \stackrel{def}{=}\left[ 
\begin{array}{cccc}
\left( a+\mathrm{i}b\right) 1_2 & 0_2 & \left( c+\mathrm{i}q\right) 1_2 & 0_2
\\ 
0_2 & 1_2 & 0_2 & 0_2 \\ 
\left( -c+\mathrm{i}q\right) 1_2 & 0_2 & \left( a-\mathrm{i}b\right) 1_2 & 
0_2 \\ 
0_2 & 0_2 & 0_2 & 1_2
\end{array}
\right] \mbox{,} 
\]

and

\begin{equation}
U^{\left( +\right) }\left( t,\mathbf{x}\right)\stackrel =\rm{R}\left( u\left(%
 t,\mathbf{x}\right),v\left( t,\mathbf{x}\right),k\left( t,\mathbf{x}\right),%
s\left( t,\mathbf{x}\right)\right)
\label{upls}
\end{equation}

with

\[
u^2\left( t,\mathbf{x}\right)+v^2\left( t,\mathbf{x}\right)+k^2\left( t,\mathbf{x}\right)%
+s^2\left( t,\mathbf{x}\right)=1 
\]

and

\begin{equation}
\mathrm{R}\left( u,v,k,s\right) \stackrel{def}{=}\left[ 
\begin{array}{cccc}
1_2 & 0_2 & 0_2 & 0_2 \\ 
0_2 & \left( u+\mathrm{i}v\right) 1_2 & 0_2 & \left( k+\mathrm{i}s\right) 1_2
\\ 
0_2 & 0_2 & 1_2 & 0_2 \\ 
0_2 & \left( -k+\mathrm{i}s\right) 1_2 & 0_2 & \left( u-\mathrm{i}v\right)
1_2
\end{array}
\right] \mbox{.}
\end{equation}

$U^{\left( +\right) }$ correspond to antileptonns since $R=S\gamma ^{\left[
5\right] }$ (\ref{xa}).

Let us consider $U^{\left( -\right) }$.

Let us denote:

\[
\ell _{\circ }\stackrel{def}{=}\imath _{\circ }\left( a,b,q,c\right) 
\mbox{,
}\ell _{*}\stackrel{def}{=}\imath _{*}\left( a,b,q,c\right) 
\]

with

\[
\imath _{\circ }\left( a,b,q,c\right) \stackrel{def}{=}\frac 1{2\sqrt{\left(
1-a^2\right) }}\left[ 
\begin{array}{cc}
\left( b+\sqrt{\left( 1-a^2\right) }\right) 1_4 & \left( q-\mathrm{i}%
c\right) 1_4 \\ 
\left( q+\mathrm{i}c\right) 1_4 & \left( \sqrt{\left( 1-a^2\right) }%
-b\right) 1_4
\end{array}
\right] 
\]

and

\[
\imath _{*}\left( a,b,q,c\right) \stackrel{def}{=}\frac 1{2\sqrt{\left(
1-a^2\right) }}\left[ 
\begin{array}{cc}
\left( \sqrt{\left( 1-a^2\right) }-b\right) 1_4 & \left( -q+\mathrm{i}%
c\right) 1_4 \\ 
\left( -q-\mathrm{i}c\right) 1_4 & \left( b+\sqrt{\left( 1-a^2\right) }%
\right) 1_4
\end{array}
\right] \mbox{.} 
\]

If

\begin{equation}
\partial _\mu U^{\left( -\right) \dagger }=U^{\left( -\right) \dagger
}\partial _\mu  \label{aa}
\end{equation}

for $\mu \in \left\{ 0,1,2,3\right\} $ then the leptonn hamiltonian is
invariant for the following global transformation:

\begin{eqnarray}
&&\widetilde{\varphi }\rightarrow \widetilde{\varphi }^{\prime }=U^{\left(
-\right) }\widetilde{\varphi }\mbox{,}  \nonumber \\
&&x_4\rightarrow x_4^{\prime }=\left( \ell _{\circ }+\ell _{*}\right)
ax_4+\left( \ell _{\circ }-\ell _{*}\right) \sqrt{1-a^2}x_5\mbox{,}
\label{glb} \\
&&x_5\rightarrow x_5^{\prime }=\left( \ell _{\circ }+\ell _{*}\right)
ax_5-\left( \ell _{\circ }-\ell _{*}\right) \sqrt{1-a^2}x_4\mbox{,} 
\nonumber \\
&&x_\mu \rightarrow x_\mu ^{\prime }=x_\mu \mbox{.}  \nonumber
\end{eqnarray}

Let (\ref{aa}) does not hold true and:

\begin{equation}
\ \ K\stackrel{def}{=}\sum_{\mu =0}^3\beta ^{\left[ \mu \right] }\left(
F_\mu +0.5g_1YB_\mu \right) \mbox{.}  \label{kdf}
\end{equation}

In that case from (\ref{gkB}) the motion equation has got the following form:

\begin{equation}
\left( K+\sum_{\mu =0}^3\beta ^{\left[ \mu \right] }\mathrm{i}\partial _\mu
+\gamma ^{\left[ 0\right] }\mathrm{i}\partial _5+\beta ^{\left[ 4\right] }%
\mathrm{i}\partial _4\right) \widetilde{\varphi }=0\mbox{.}  \label{me81}
\end{equation}

Hence for the following transformations:

\begin{eqnarray}
&&\widetilde{\varphi }\rightarrow \widetilde{\varphi }^{\prime }\stackrel{def%
}{=}U^{\left( -\right) }\widetilde{\varphi }\mbox{,}  \nonumber \\
&&x_4\rightarrow x_4^{\prime }\stackrel{def}{=}\left( \ell _{\circ }+\ell
_{*}\right) ax_4+\left( \ell _{\circ }-\ell _{*}\right) \sqrt{1-a^2}x_5%
\mbox{,}  \nonumber \\
&&x_5\rightarrow x_5^{\prime }\stackrel{def}{=}\left( \ell _{\circ }+\ell
_{*}\right) ax_5-\left( \ell _{\circ }-\ell _{*}\right) \sqrt{1-a^2}x_4%
\mbox{,}  \label{gll} \\
&&x_\mu \rightarrow x_\mu ^{\prime }\stackrel{def}{=}x_\mu \mbox{, for }\mu
\in \left\{ 0,1,2,3\right\} \mbox{,}  \nonumber \\
&&K\rightarrow K^{\prime }  \nonumber
\end{eqnarray}

with

\[
\begin{array}{c}
\partial _4U^{\left( -\right) }=U^{\left( -\right) }\partial _4\mbox{ and }%
\partial _5U^{\left( -\right) }=U^{\left( -\right) }\partial _5
\end{array}
\]

this equation has got the following form (see Appendix):

\begin{equation}
\left( 
\begin{array}{c}
U^{\left( -\right) \dagger }K^{\prime }U^{\left( -\right) }+ \\ 
+\sum_{\mu =0}^3\beta ^{\left[ \mu \right] }\mathrm{i}\left( \partial _\mu
+U^{\left( -\right) \dagger }\left( \partial _\mu U^{\left( -\right)
}\right) \right) +\gamma ^{\left[ 0\right] }\mathrm{i}\partial _5+\beta
^{\left[ 4\right] }\mathrm{i}\partial _4
\end{array}
\right) \widetilde{\varphi }=0\mbox{.}  \label{me82}
\end{equation}

Therefore if

\begin{equation}
K^{\prime }=K-\mathrm{i}\sum_{\mu =0}^3\beta ^{\left[ \mu \right] }\left(
\partial _\mu U^{\left( -\right) }\right) U^{\left( -\right) \dagger }
\label{ksht}
\end{equation}

then the equation (\ref{me81}) is invariant for the local transformation (%
\ref{gll}).

Let $g_2$ be some positive real number.

If design ($a,b,c,q$ form $U^{\left( -\right) }$):

\[
\begin{array}{c}
W_\mu ^{0,}\stackrel{def}{=}-\frac 2{g_2q}\left( 
\begin{array}{c}
q\left( \partial _\mu a\right) b-q\left( \partial _\mu b\right) a+\left(
\partial _\mu c\right) q^2+ \\ 
+a\left( \partial _\mu a\right) c+b\left( \partial _\mu b\right) c+c^2\left(
\partial _\mu c\right)
\end{array}
\right) \\ 
W_\mu ^{1,}\stackrel{def}{=}-\frac 2{g_2q}\left( 
\begin{array}{c}
\left( \partial _\mu a\right) a^2-bq\left( \partial _\mu c\right) +a\left(
\partial _\mu b\right) b+ \\ 
+a\left( \partial _\mu c\right) c+q^2\left( \partial _\mu a\right) +c\left(
\partial _\mu b\right) q
\end{array}
\right) \\ 
W_\mu ^{2,}\stackrel{def}{=}-\frac 2{g_2q}\left( 
\begin{array}{c}
q\left( \partial _\mu a\right) c-a\left( \partial _\mu a\right) b-b^2\left(
\partial _\mu b\right) - \\ 
-c\left( \partial _\mu c\right) b-\left( \partial _\mu b\right) q^2-\left(
\partial _\mu c\right) qa
\end{array}
\right)
\end{array}
\]

and

\[
W_\mu \stackrel{def}{=}\left[ 
\begin{array}{cccc}
W_\mu ^{0,}1_2 & 0_2 & \left( W_\mu ^{1,}-\mathrm{i}W_\mu ^{2,}\right) 1_2 & 
0_2 \\ 
0_2 & 0_2 & 0_2 & 0_2 \\ 
\left( W_\mu ^{1,}+\mathrm{i}W_\mu ^{2,}\right) 1_2 & 0_2 & -W_\mu ^{0,}1_2
& 0_2 \\ 
0_2 & 0_2 & 0_2 & 0_2
\end{array}
\right] 
\]

then

\begin{equation}
-\mathrm{i}\left( \partial _\mu U^{\left( -\right) }\right) U^{\left(
-\right) \dagger }=\frac 12g_2W_\mu \mbox{,}  \label{w}
\end{equation}

and from (\ref{w}), (\ref{kdf}), (\ref{ksht}), (\ref{me81}):

\begin{equation}
\left( 
\begin{array}{c}
\sum_{\mu =0}^3\beta ^{\left[ \mu \right] }\mathrm{i}\left( \partial _\mu -%
\mathrm{i}0.5g_1B_\mu Y-\mathrm{i}\frac 12g_2W_\mu -\mathrm{i}F_\mu \right)
\\ 
+\gamma ^{\left[ 0\right] }\mathrm{i}\partial _5^{\prime }+\beta ^{\left[
4\right] }\mathrm{i}\partial _4^{\prime }
\end{array}
\right) \widetilde{\varphi }^{\prime }=0\mbox{.}  \label{hW}
\end{equation}

Let

\[
U^{\prime }\stackrel{def}{=}\mathrm{S}\left( a^{\prime },b^{\prime
},c^{\prime },q^{\prime }\right) \mbox{.} 
\]

In this case if

\[
U^{\prime \prime }\stackrel{def}{=}U^{\prime }U^{\left( -\right) } 
\]

then there exist some real functions $a^{\prime \prime }\left( t,\mathbf{x}%
\right) $, $b^{\prime \prime }\left( t,\mathbf{x}\right) $, $c^{\prime
\prime }\left( t,\mathbf{x}\right) $, $q^{\prime \prime }\left( t,\mathbf{x}%
\right) $ such that $U^{\prime \prime }$ has got the similar form:

\[
U^{\prime \prime }\stackrel{def}{=}\mathrm{S}\left( a^{\prime \prime
},b^{\prime \prime },c^{\prime \prime },q^{\prime \prime }\right) \mbox{.} 
\]

If

\[
\ell _{\circ }^{\prime \prime }\stackrel{def}{=}\imath _{\circ }\left(
a^{\prime \prime },b^{\prime \prime },q^{\prime \prime },c^{\prime \prime
}\right) \mbox{, }\ell _{*}^{\prime \prime }\stackrel{def}{=}\imath
_{*}\left( a^{\prime \prime },b^{\prime \prime },q^{\prime \prime
},c^{\prime \prime }\right) \mbox{;} 
\]

\begin{eqnarray}
&&\widetilde{\varphi }\rightarrow \widetilde{\varphi }^{\prime \prime }%
\stackrel{def}{=}U^{\prime \prime }\widetilde{\varphi }\mbox{,}  \nonumber \\
&&x_4\rightarrow x_4^{\prime \prime }\stackrel{def}{=}\left( \ell _{\circ
}^{\prime \prime }+\ell _{*}^{\prime \prime }\right) a^{\prime \prime
}x_4+\left( \ell _{\circ }^{\prime \prime }-\ell _{*}^{\prime \prime
}\right) \sqrt{1-a^{\prime \prime 2}}x_5\mbox{,}  \nonumber \\
&&x_5\rightarrow x_5^{\prime \prime }\stackrel{def}{=}\left( \ell _{\circ
}^{\prime \prime }+\ell _{*}^{\prime \prime }\right) a^{\prime \prime
}x_5-\left( \ell _{\circ }^{\prime \prime }-\ell _{*}^{\prime \prime
}\right) \sqrt{1-a^{\prime \prime 2}}x_4\mbox{,}  \label{tt2} \\
&&x_\mu \rightarrow x_\mu ^{\prime \prime }\stackrel{def}{=}x_\mu 
\mbox{,
for }\mu \in \left\{ 0,1,2,3\right\} \mbox{,}  \nonumber \\
&&K\rightarrow K^{\prime \prime }\stackrel{def}{=}\sum_{\mu =0}^3\beta
^{\left[ \mu \right] }\left( F_\mu +0.5g_1YB_\mu +\frac 12g_2W_\mu ^{\prime
\prime }\right)  \nonumber
\end{eqnarray}

then from (\ref{w}):

\[
W_\mu ^{\prime \prime }=-\frac{2i}{g_2}\left( \partial _\mu \left( U^{\prime
}U^{\left( -\right) }\right) \right) \left( U^{\prime }U^{\left( -\right)
}\right) ^{\dagger }\mbox{.} 
\]

Hence:

\[
W_\mu ^{\prime \prime }=-\frac{2i}{g_2}\left( \partial _\mu U^{\prime
}\right) U^{\prime \dagger }-\frac{2i}{g_2}U^{\prime }\left( \partial _\mu
U^{\left( -\right) }\right) U^{\left( -\right) \dagger }U^{\prime \dagger }%
\mbox{,} 
\]

i.e. from (\ref{w}):

\begin{equation}
W_\mu ^{\prime \prime }=U^{\prime }W_\mu U^{\prime \dagger }-\frac{2i}{g_2}%
\left( \partial _\mu U^{\prime }\right) U^{\prime \dagger }  \label{w00}
\end{equation}

as in Standard Model.

The motion equation of the Yang-Mills SU(2) field at the space without
matter (for instance \cite{Sd} or \cite{Rd} ) has got the following form:

\[
\partial ^\nu \mathbf{W}_{\mu \nu }=-g_2\mathbf{W}^\nu \times \mathbf{W}%
_{\mu \nu } 
\]

with:

\[
\mathbf{W}_{\mu \nu }=\partial _\mu \mathbf{W}_\nu -\partial _\nu \mathbf{W}%
_\mu +g_2\mathbf{W}_\mu \times \mathbf{W}_\nu 
\]

and

\[
\mathbf{W}_\mu =\left[ 
\begin{array}{c}
W_\mu ^{0,} \\ 
W_\mu ^{1,} \\ 
W_\mu ^{2,}
\end{array}
\right] \mbox{.} 
\]

Hence the motion equation for $W_\mu ^{0,}$ is the following:

\begin{equation}
\begin{array}{c}
\partial ^\nu \partial _\nu W_\mu ^{0,}=g_2^2\left( W^{2,\nu }W_\nu
^{2,}+W^{1,\nu }W_\nu ^{1,}\right) W_\mu ^{0,}- \\ 
-g_2^2\left( W^{1,\nu }W_\mu ^{1,}+W^{2,\nu }W_\mu ^{2,}\right) W_\nu ^{0,}+
\\ 
+g_2\partial ^\nu \left( W_\mu ^{1,}W_\nu ^{2,}-W_\mu ^{2,}W_\nu
^{1,}\right) + \\ 
+g_2\left( W^{1,\nu }\partial _\mu W_\nu ^{2,}-W^{1,\nu }\partial _\nu W_\mu
^{2,}-W^{2,\nu }\partial _\mu W_\nu ^{1,}+W^{2,\nu }\partial _\nu W_\mu
^{1,}\right) + \\ 
+\partial ^\nu \partial _\mu W_\nu ^{0,}
\end{array}
\label{b}
\end{equation}

with $g_{0,0}=1$, $g_{1,1}=g_{2,2}=g_{3,3}=-1$ (i.e.: $W^\nu W_\nu
=W^0W_0-W^1W_1-W^2W_2-W^3W_3$. For the gauging with $W_0=0$: $W^\nu W_\nu
=-\left( W^1W_1+W^2W_2+W^3W_3\right) $ ).

$W_\mu ^{1,}$ and $W_\mu ^{2,}$ satisfy to similar equations.

This equation can be reformed as the following:

\[
\begin{array}{c}
\partial ^\nu \partial _\nu W_\mu ^{0,}=\left[ g_2^2\left( W^{2,\nu }W_\nu
^{2,}+W^{1,\nu }W_\nu ^{1,}+W^{0,\nu }W_\nu ^{0,}\right) \right] \cdot W_\mu
^{0,}- \\ 
-g_2^2\left( W^{1,\nu }W_\mu ^{1,}+W^{2,\nu }W_\mu ^{2,}+W^{0,\nu }W_\mu
^{0,}\right) W_\nu ^{0,}+ \\ 
+g_2\partial ^\nu \left( W_\mu ^{1,}W_\nu ^{2,}-W_\mu ^{2,}W_\nu
^{1,}\right) + \\ 
+g_2\left( W^{1,\nu }\partial _\mu W_\nu ^{2,}-W^{1,\nu }\partial _\nu W_\mu
^{2,}-W^{2,\nu }\partial _\mu W_\nu ^{1,}+W^{2,\nu }\partial _\nu W_\mu
^{1,}\right) + \\ 
+\partial ^\nu \partial _\mu W_\nu ^{0,}\mbox{.}
\end{array}
\]

This equation looks like to the Klein-Gordon equation of field $W_\mu ^{0,}$
with mass

\begin{equation}
g_2\left[ -\left( W^{2,\nu }W_\nu ^{2,}+W^{1,\nu }W_\nu ^{1,}+W^{0,\nu
}W_\nu ^{0,}\right) \right] ^{\frac 12}.  \label{z10}
\end{equation}

and with the additional terms of the $W_\mu ^{0,}$ interactions with others
components of $\mathbf{W}$.

"Mass" (\ref{z10}) is invariant for the following transformations:

\[
\left\{ 
\begin{array}{c}
W_r^{k,\prime }=W_r^{k,}\cos \lambda -W_s^{k,}\sin \lambda \mbox{,} \\ 
W_s^{k,\prime }=W_r^{k,}\sin \lambda +W_s^{k,}\cos \lambda \mbox{;}
\end{array}
\right| 
\]

\[
\left\{ 
\begin{array}{c}
W_0^{k,\prime }=W_0^{k,}\cosh \lambda -W_s^{k,}\sinh \lambda \mbox{,} \\ 
W_s^{k,\prime }=W_s^{k,}\cosh \lambda -W_0^{k,}\sinh \lambda
\end{array}
\right| 
\]

with real number $\lambda $ and $r\in \left\{ 1,2,3\right\} $ and $s\in
\left\{ 1,2,3\right\} $, and (\ref{z10}) is invariant for global weak
isospin transformation $U^{\prime }$:

\[
W_\nu ^{\prime }\rightarrow W_\nu ^{\prime \prime }=U^{\prime }W_\nu
U^{\prime \dagger } 
\]

but is not invariant for local transformation (\ref{w00})

Equation (\ref{b}) can be simplified as follows:

\[
\begin{array}{c}
\sum_\nu g_{\nu ,\nu }\partial ^\nu \partial _\nu W_\mu ^{0,}=\left[
g_2^2\sum_{\nu \neq \mu }g_{\nu ,\nu }\left( \left( W_\nu ^{2,}\right)
^2+\left( W_\nu ^{1,}\right) ^2\right) \right] \cdot W_\mu ^{0,}- \\ 
-g_2^2\sum_{\nu \neq \mu }g_{\nu ,\nu }\left( W^{1,\nu }W_\mu ^{1,}+W^{2,\nu
}W_\mu ^{2,}\right) W_\nu ^{0,}- \\ 
+g_2\sum_\nu g_{\nu ,\nu }\partial ^\nu \left( W_\mu ^{1,}W_\nu ^{2,}-W_\mu
^{2,}W_\nu ^{1,}\right) + \\ 
+g_2\sum_\nu g_{\nu ,\nu }\left( W^{1,\nu }\partial _\mu W_\nu
^{2,}-W^{1,\nu }\partial _\nu W_\mu ^{2,}-W^{2,\nu }\partial _\mu W_\nu
^{1,}+W^{2,\nu }\partial _\nu W_\mu ^{1,}\right) + \\ 
+\partial _\mu \sum_\nu g_{\nu ,\nu }\partial ^\nu W_\nu ^{0,}\mbox{.}
\end{array}
\]

(here no of summation over indexes $_\nu ^\nu $; the summation is expressed
by $\sum $ ) .

In this equation the form

\[
g_2\left[ -\sum_{\nu \neq \mu }g_{\nu ,\nu }\left( \left( W_\nu ^{2,}\right)
^2+\left( W_\nu ^{1,}\right) ^2\right) \right] ^{\frac 12} 
\]

varies at space, but this does not contain $W_\mu ^{0,}$ and locally acts as
mass, i.e. this does not allow to particles of this field to behave as a
massless ones.

Let

\[
\begin{array}{c}
\alpha \stackrel{def}{=}\arctan \frac{g_1}{g_2}\mbox{,} \\ 
Z_\mu \stackrel{def}{=}\left( W_\mu ^{0,}\cos \alpha -B_\mu \sin \alpha
\right) \mbox{,} \\ 
A_\mu \stackrel{def}{=}\left( B_\mu \cos \alpha +W_\mu ^{0,}\sin \alpha
\right) \mbox{.}
\end{array}
\]

In that case:

\[
\begin{array}{c}
\sum_\nu g_{\nu ,\nu }\partial _\nu \partial _\nu W_\mu ^{0,}=\cos \alpha
\cdot \sum_\nu g_{\nu ,\nu }\partial _\nu \partial _\nu Z_\mu +\sin \alpha
\cdot \sum_\nu g_{\nu ,\nu }\partial _\nu \partial _\nu A_\mu \mbox{.}
\end{array}
\]

If

\[
\sum_\nu g_{\nu ,\nu }\partial _\nu \partial _\nu A_\mu =0 
\]

then

\[
m_Z=\frac{m_W}{\cos \alpha } 
\]

with $m_W$ from (\ref{z10}). This is almost as in Standard Model.

\section{Conclusion}

Therefore we ourselves choose the expression of a probability in the form (%
\ref{ro}). We ourselves introduce a creation operator and an annihilation
operator of an event probability. We ourselves add two more quasispace
coordinates to our three and etc. That is we construct the logical structure
such that there are inevitable particles, antiparticles and gauge bosons.
Sort of answer is defined by sort of question.

Thus we ourselves make up the rules of the probabilistic information
processing in the form of the laws of the quantum theory. And the values of
parameters are depend on the structure of our devices.

This is likely that Universe gives only probabilities of events. But the
 physics laws, operating these probabilities, are the result of our device
structure properties and of the logic behavior of our language. If the other
methods of the information receiving and processing can exist somewhere then
the physical laws of the other shape must operate there. Thus the quantum
theory is only one amongst possible ways for the processing of a probabilistic 
information.

%\bigskip

%{\bf Acknowledgment}

%\bigskip

%The large thanks to Prof. V. Dvoeglazov.

%\bigskip
\section{Appendix. Operations at space $\Im _{\jmath }$}

Let $\widetilde{\varphi }\in \Im _{\jmath }$. That is:

\[
\widetilde{\varphi }\left( t,\mathbf{x},x_5,x_4\right) =\left( 
\begin{array}{c}
\exp \left( -\mathrm{i}hs_0x_4\right) \sum_{r=1}^4\phi _r\left( t,\mathbf{x}%
,0,s\right) \epsilon _r \\ 
+\exp \left( -\mathrm{i}hn_0x_5\right) \sum_{j=5}^8\phi _j\left( t,\mathbf{x}%
,n,0\right) \epsilon _j
\end{array}
\right) \mbox{.} 
\]

Let linear operators $\beta $ and $K$ act in the basis $\epsilon _k$ and the
linear operator $U^{\left( -\right) }$ acts at space $\Im _{\jmath }$.

In this case if

\[
U^{\left( -\right) }=\left[ 
\begin{array}{cc}
U_{1,1} & U_{1,2} \\ 
U_{2,1} & U_{2,2}
\end{array}
\right] 
\]

at basis $\jmath $ with $U_{j,r}$ as 4$\times $4 matrices then $U_{j,r}$ act
at basis $\epsilon _k$, too.

Example of calculation:

$\left( K-\beta U^{\left( -\right) }\right) \widetilde{\varphi }=$

$=\left( K-\beta U^{\left( -\right) }\right) \left( 
\begin{array}{c}
\exp \left( -\mathrm{i}hs_0x_4\right) \sum_{r=1}^4\phi _r\left( t,\mathbf{x}%
,0,s\right) \epsilon _r \\ 
+\exp \left( -\mathrm{i}hn_0x_5\right) \sum_{j=5}^8\phi _j\left( t,\mathbf{x}%
,n,0\right) \epsilon _j
\end{array}
\right) $

$=\left( K-\beta U^{\left( -\right) }\right) \exp \left( -\mathrm{i}%
hs_0x_4\right) \sum_{r=1}^4\phi _r\left( t,\mathbf{x},0,s\right) \epsilon _r$

$+\left( K-\beta U^{\left( -\right) }\right) \exp \left( -\mathrm{i}%
hn_0x_5\right) \sum_{j=5}^8\phi _j\left( t,\mathbf{x},n,0\right) \epsilon _j$

$=K\exp \left( -\mathrm{i}hs_0x_4\right) \sum_{r=1}^4\phi _r\left( t,\mathbf{%
x},0,s\right) \epsilon _r$

$-\beta U\exp \left( -\mathrm{i}hs_0x_4\right) \sum_{r=1}^4\phi _r\left( t,%
\mathbf{x},0,s\right) \epsilon _r$

$+K\exp \left( -\mathrm{i}hn_0x_5\right) \sum_{j=5}^8\phi _j\left( t,\mathbf{%
x},n,0\right) \epsilon _j$

$-\beta U\exp \left( -\mathrm{i}hn_0x_5\right) \sum_{j=5}^8\phi _j\left( t,%
\mathbf{x},n,0\right) \epsilon _j$

$=\exp \left( -\mathrm{i}hs_0x_4\right) \sum_{r=1}^4\phi _r\left( t,\mathbf{x%
},0,s\right) K\epsilon _r$

$-\beta \exp \left( -\mathrm{i}hs_0x_4\right) \left( 
\begin{array}{c}
U_{1,1}\sum_{r=1}^4\phi _r\left( t,\mathbf{x},0,s\right) \epsilon _r \\ 
+U_{1,2}\sum_{j=5}^8\phi _j\left( t,\mathbf{x},n,0\right) \epsilon _j
\end{array}
\right) $

$+\exp \left( -\mathrm{i}hn_0x_5\right) \sum_{j=5}^8\phi _j\left( t,\mathbf{x%
},n,0\right) K\epsilon _j$

$-\beta \exp \left( -\mathrm{i}hn_0x_5\right) \left( 
\begin{array}{c}
U_{2,1}\sum_{r=1}^4\phi _r\left( t,\mathbf{x},0,s\right) \epsilon _r \\ 
+U_{2,2}\sum_{j=5}^8\phi _j\left( t,\mathbf{x},n,0\right) \epsilon _j
\end{array}
\right) $

$=\exp \left( -\mathrm{i}hs_0x_4\right) \sum_{r=1}^4\phi _r\left( t,\mathbf{x%
},0,s\right) K\epsilon _r$

$-\beta \exp \left( -\mathrm{i}hs_0x_4\right) \left( 
\begin{array}{c}
\sum_{r=1}^4\phi _r\left( t,\mathbf{x},0,s\right) U_{1,1}\epsilon _r \\ 
+\sum_{j=5}^8\phi _j\left( t,\mathbf{x},n,0\right) U_{1,2}\epsilon _j
\end{array}
\right) $

$+\exp \left( -\mathrm{i}hn_0x_5\right) \sum_{j=5}^8\phi _j\left( t,\mathbf{x%
},n,0\right) K\epsilon _j$

$-\beta \exp \left( -\mathrm{i}hn_0x_5\right) \left( 
\begin{array}{c}
\sum_{r=1}^4\phi _r\left( t,\mathbf{x},0,s\right) U_{2,1}\epsilon _r \\ 
+\sum_{j=5}^8\phi _j\left( t,\mathbf{x},n,0\right) U_{2,2}\epsilon _j
\end{array}
\right) $

$=\exp \left( -\mathrm{i}hs_0x_4\right) \sum_{r=1}^4\phi _r\left( t,\mathbf{x%
},0,s\right) K\epsilon _r$

$-\exp \left( -\mathrm{i}hs_0x_4\right) \left( 
\begin{array}{c}
\sum_{r=1}^4\phi _r\left( t,\mathbf{x},0,s\right) \beta U_{1,1}\epsilon _r
\\ 
+\sum_{j=5}^8\phi _j\left( t,\mathbf{x},n,0\right) \beta U_{1,2}\epsilon _j
\end{array}
\right) $

$+\exp \left( -\mathrm{i}hn_0x_5\right) \sum_{j=5}^8\phi _j\left( t,\mathbf{x%
},n,0\right) K\epsilon _j$

$-\exp \left( -\mathrm{i}hn_0x_5\right) \left( 
\begin{array}{c}
\sum_{r=1}^4\phi _r\left( t,\mathbf{x},0,s\right) \beta U_{2,1}\epsilon _r
\\ 
+\sum_{j=5}^8\phi _j\left( t,\mathbf{x},n,0\right) \beta U_{2,2}\epsilon _j
\end{array}
\right) $.

\vskip 30pt
\begin{fref}

\bibitem{Md}  for instance, E. Madelung, {\it Die Mathematischen
Hilfsmittel des Physikers.} (Springer Verlag, 1957) p.29

\bibitem{Q42}  G. A. Quznetsov, Physical events and quantum field theory
without Higgs, {\it preprint} hep-ph/9812339, 2004, p.42

\bibitem{D5}  For examples: A. Soddi, N-K. Tran, Democratic Mass Matrices
From Five Dimensions, hep-ph/0308043; D. Chang, Ch-Sh. Chen, Ch-H. Chou, H.
Hatanaka, A model of CP violation from extra dimension, hep-ph/0406059;
Janusz Garecki, On Hypothesis of the two large extradimensions,
gr-qc/0306041; C. Csaki, C. Grojean, H. Murayama, L. Pilo, J. Terning, Gauge
theories on interval: Unitarity without a Higgs, hep-ph/0305237; H.
Davoudiasl, J. L. Hewett, B. Lillie, T. G. Rizzo, Higgless electroweak
simmetry breaking in wraped backgrounds: Constraints and signatures,
hep-ph/0312193; S. Gabriel, S. Nandi, G. Seidl, 6D Higgless Standard Model,
hep-ph/0406020; T. Rizzo, Phenomenology of Higgsless electroweak symmetry
breaking, hep-ph/0405094

\bibitem{VVD} V. V. Dvoeglazov, Fizika, \textbf{B6}, No. 3, pp. 111-122 (1997). 
Int. J. Theor. Phys., \textbf{34}, No. 12, pp. 2467-2490 (1995).
Annales de la Fondation de Louis de Broglie, \textbf{25}, No. 1, pp. 81-91 (2000).   

\bibitem{Barut} A. O. Barut, P. Cordero, G. C. Ghirardi, Nuovo. Cim. 
\textbf{A66}, 36 (1970). A. O. Barut, Phys. Let. \textbf{73B}, 310 (1978); 
Phys. Rev. Let. \textbf{42}, 1251 (1979). A. O. Barut, P. Cordero, 
G. C. Ghirardi, Phys. Rev. \textbf{182}, 1844 (1969).

\bibitem{Wilson} R. Wilson, Nucl. Phys. \textbf{B68}, 157 (1974).

\bibitem{DVB} V. V. Dvoeglazov, Additional Equations Derived from the Ryder
Postulates in the (1/2,0)+(0,1/2) Representation of the Lorentz Group.
hep-th/9906083. Int. J. Theor. Phys. {\bf 37} (1998) 1909. Helv. Phys.
Acta {\bf 70} (1997) 677. Fizika B {\bf 6} (1997) 75; Int. J. Theor.
Phys. {\bf 34} (1995) 2467. Nuovo Cimento {\bf 108} A (1995) 1467.
Nuovo Cimento {\bf 111} B (1996) 483. Int. J. Theor. Phys. {\bf 36}
(1997) 635.

\bibitem{AV}  D. V. Ahluwalia, (j,0)+(0,j) Covariant spinors and causal
propagators based on Weinberg formalism. nucl-th/9905047. Int. J. Mod. Phys.
A {\bf 11} (1996) 1855.

\bibitem{Sd}  M. V. Sadovski, {\it Lectures on the Quantum Fields Theory}
(Institute of Electrophysics, UrO RAS, 2000), p.33 (2.12)

\bibitem{Rd}  Lewis H. Ryder, {\it Quantum Field Theory}, Mir, Moscow
(1987), p.133, (3.136)

% eref si anglais
\end{fref}

{\it \'Electromagn\'etiques calibriques champs, les particules et antiparticules 
prennent naissance des probabilit\'es}

\man{26 avril 2004}
\end{document}